\documentclass[journal,twoside,web]{ieeecolor}

\usepackage{generic}
\usepackage{cite}
\usepackage{amsmath,amssymb,amsfonts}
\usepackage{algorithmic}
\usepackage{graphicx}
\usepackage{textcomp}

\usepackage{graphicx}
\usepackage{subfigure}
\usepackage{epsfig} 
\usepackage{cite}
\usepackage{lcsys}

\usepackage{cite}
\usepackage{amsmath,amssymb,amsfonts}
\usepackage{algorithmic}
\usepackage{graphicx}
\usepackage{textcomp}
\usepackage{tabularx,booktabs,textcomp}
\usepackage{algorithm}
\usepackage{hyperref}
\usepackage{xcolor}
\usepackage{array}
\usepackage{textcomp}
\usepackage{setspace}
\usepackage{stfloats}
\usepackage{xurl}
\usepackage{url}

\usepackage[normalem]{ulem}
\usepackage{amsthm}
\usepackage{verbatim}
\usepackage{graphicx}
\usepackage[english]{babel}
\newtheorem{thm}{Theorem}

\newtheorem{remark}{Remark}

\DeclareMathOperator*{\argmin}{arg\,min}
\usepackage{amsmath,amssymb,amsthm}

\begin{document}

\def\BibTeX{{\rm B\kern-.05em{\sc i\kern-.025em b}\kern-.08em
    T\kern-.1667em\lower.7ex\hbox{E}\kern-.125emX}}
\markboth{\journalname, VOL. XX, NO. XX, XXXX 2017}
{Author \MakeLowercase{\textit{et al.}}: Preparation of Papers for IEEE Control Systems Letters (August 2022)}

\title{Optimal Storage and Solar Capacity of a Residential Household under Net Metering and Time-of-Use Pricing}
\author{{Victor Sam Moses Babu K.$^{a}$,~\IEEEmembership{Member,~IEEE}, Pratyush Chakraborty$^{a}$,~\IEEEmembership{Member,~IEEE}, Enrique Baeyens$^{b}$,~\IEEEmembership{Senior Member,~IEEE}, and Pramod P. Khargonekar$^{c}$,~\IEEEmembership{Fellow,~IEEE}}
\thanks{$^{a}$K. Victor Sam Moses Babu and Pratyush Chakraborty are with the Department of Electrical and Electronics Engineering, BITS Pilani Hyderabad Campus, Hyderabad 500078, India (e-mail: victorsam.k@gmail.com, pchakraborty@hyderabad.bits-pilani.ac.in).}
\thanks{$^{b}$E. Baeyens is with Instituto de las Tecnologías Avanzadas de la Producción, Universidad de Valladolid, 47011 Valladolid, Spain (e-mail: enrbae@eis.uva.es).}
\thanks{$^{c}$P. Khargonekar is with the Department of Electrical Engineering and Computer Science, University of California, Irvine, CA 92697 USA (e-mail: pramod.khargonekar@uci.edu).}
\thanks{We acknowledge the contribution of Satya Surya Vinay K, a student of BITS Pilani, Hyderabad Campus for helping us in checking the calculations of the paper. }}

\maketitle
\thispagestyle{empty}

\begin{abstract}
Incentive programs and ongoing reduction in costs are driving joint installation of solar PV panels and storage systems in residential households. There is a need for optimal investment decisions to reduce the electricity consumption costs of the households further. In this paper, we first develop analytical expression of storage investment decision and then of solar investment decision for a household which is under net metering billing mechanism with time of use pricing condition. Using real data of a residential household in Austin, TX, USA, we study how the investment decisions would provide benefit for a period of one year. Results show significant profit when using storage devices and solar panels optimally for the system. It is important to note that though our approach can help significantly to take investment decisions, the solution will still be sub-optimal for somebody who needs optimal investment jointly on both storage and solar systems. 
\end{abstract}

\begin{IEEEkeywords}
Optimal investment, time-of-use pricing, net metering, electricity storage, solar PV generation, distributed generation.
\end{IEEEkeywords}


\section{Introduction}
\label{section:Introduction}
Investment in energy storage devices and solar PV panels for household utilization has recently increased substantially. The United States installed a record 11.8 GW of solar PV and  5.7 GWh of storage units within the first nine months of 2021, with around 1 GW of residential PV and 250 MWh of residential storage installed in every quarter \cite{1}. The use of energy storage devices with solar PV panels allows for better use of solar energy. But as of December 2020, only 30\% of battery storage systems were installed with other renewable energy resources like wind or solar \cite{2}. We expect the relationship between solar and storage to change significantly in the USA over the next few years as many upcoming projects plan to make 80\% of the battery storage power capacity to be paired with solar operational between 2021 and 2023 \cite{1}. Residential installations of battery units are eligible for investment tax credit if paired with solar \cite{FTI}. Also, proposed national and state policies like the California state policy are intended to incentivize the use of PV with storage \cite{1}. Storage enables increased self-consumption in solar PV and also allows for allocating the excess solar energy to different time-of-use rate periods.  Power injection from energy storage devices into the grid is also becoming popular with more and more countries implementing policies for encouraging self-consumption or demand-side management and its incorporation into electricity markets through aggregators or energy communities to improve efficiency \cite{28},\cite{33}.
The capital costs for solar PV and energy storage continue to reduce, and it is observed that there is a reduction from 8\% to 13\% in 2021 from the previous year \cite{1}. Major companies have turned towards combined PV and storage installations. Sunrun has installed 28,000 units of PV with storage systems. In the third quarter of 2021, SunPower reported that 27\% of its solar customers had purchased battery storage directly through the company \cite{1}. Tesla announced in April 2021 that it would only sell solar paired with storage \cite{1,2}.

Three main billing mechanisms enabling prosumers to sell their excess generated power back to the grid have been used: feed-in tariff (FiT), net metering (NM) and net purchase and sale (NPS) \cite{3}. In the FiT program, the household must sell the entire generation at a selling price and buy the total consumption from the grid at retail price. The generation and consumption are compared moment by moment in NPS program while in the NM program they are compared at the end of a billing period \cite{4}. In all these programs, we have one price for buying electricity supply from the grid and another price for selling excess energy to the grid. The net metering program is the most widely used billing program. Now with time-of-use (ToU) tariff policy included in the net metering program, there are two or more prices for buying and selling for two or more peak and off-peak periods \cite{NM2}, \cite{34}. With this innovative pricing strategy, consumers can now alter their energy usage based on the ToU pricing conditions to reduce their electricity bills. Also, ToU tariffs unlock demand-side flexibility and help in the increase of renewable energy installations \cite{5}. By combining solar PV panels and energy storage devices, we can use ToU policies in net metering programs to our advantage by storing the solar energy in storage devices during off-peak periods and selling back to the grid during peak-periods.

A key question is: how should the capacity of battery storage system be optimized in joint installations with PV at the residential level? In \cite{30,31}, the sizing of energy storage systems is considered for uncertain conditions of solar and wind energy using a model predictive control approach.  In \cite{6}, an optimal battery storage system sizing for solar-plus-storage and wind-plus-storage systems was proposed so that investors can determine the storage capacity that maximizes profitability. In \cite{7}, a cost-efficient sizing strategy of an energy storage unit showed advantages in effective utilization of investment in terms of payback time and rate of return. Xiang et al. \cite{8} proposed optimal sizing of energy storage system in active distribution networks. A comprehensive storage optimal sizing solution for microgrid applications was proposed in \cite{9}, which also determined the reliability of the microgrid. Combining storage units in a grid-integrated household becomes a beneficial option, but if solar PV is not selected optimally, it may not offer economic benefits \cite{10,11}. Thus, determining the optimal capacity of PV and storage is an essential problem for a grid-integrated household to attain maximal cost-benefit \cite{12,13}. A new optimal design approach for PV-battery microgrids was proposed in \cite{15} that calculates the optimal number of PV panels and the optimal value of the battery bank and minimizes the levelized cost of energy for a residential microgrid. Chatterji et al. \cite{16} presented a model with a customer-centric optimization by considering net metering policy and time-of-use grid pricing. But the optimal cost of investing in storage and PV was not considered. In \cite{17}, an enhanced technique for optimizing the capacity of a residential microgrid with renewable energy sources like solar and wind with storage units was presented. Hafiz et al. \cite{18} have considered ToU pricing for a solar PV generation with battery energy storage but have discouraged selling back to the grid. Zhang et al. \cite{19} analyzed whether a household with solar PV should invest in a storage unit, and participate in a FiT or NM program. In \cite{20}, the sizing of PV with energy storage system for a NM program was formulated. The investment decision in optimal storage with ToU pricing was considered \cite{21}; however, solar investment for the household was not presented. Optimal solar investment for a NM mechanism was discussed \cite{22}, while the ToU pricing and storage investment were not evaluated. The optimal solar PV and storage capacity for grid-connected households was determined for two different configurations: only PV and PV with storage for flat retail price of electricity \cite{23}. The optimal capacity of PV and storage for ToU pricing for a household was discussed, but the net metering mechanism was not considered \cite{24}.

From the above literature, it is clear that while different ToU pricing mechanisms were presented in several works, net metering is not discussed along with ToU pricing. Also, the optimal investment decisions for ToU pricing and net metering mechanisms are not evaluated thus far for a household with both solar PV and storage system. Therefore, this work focuses on how the optimal investment in solar and storage units for a grid-connected residential system would lead to a reduction in electricity consumption costs over an extended period of time. We investigate the household cost in a time-of-use setup with a net metering billing mechanism. Significant contributions of this work include:

1) We derive an expression to compute the optimal capacity for energy storage in a household with net metering and time-of-use pricing.

2) We then consider the household to have invested in this optimal storage capacity and we derive an expression to compute the optimal capacity for solar PV panels.

3) We analyse how the optimal investments in storage and solar PV capacity impact the electricity costs of the household using real data for a period of one year. The results show that significant cost savings can be obtained.




The rest of the paper is organized as follows. Section \ref{section:Problem Formulation and Results} presents the mathematical formulation and results of the proposed model. In Section \ref{section:Simulation Study and Result Analysis}, an analysis of the model with real-world data is presented. Finally, conclusions are drawn in Section \ref{section:Conclusion}.

\section{Problem Formulation and Results}
\label{section:Problem Formulation and Results}
\subsection{Model description}
\begin{figure}[ht]
  \centering
  \includegraphics[width=2.35in]{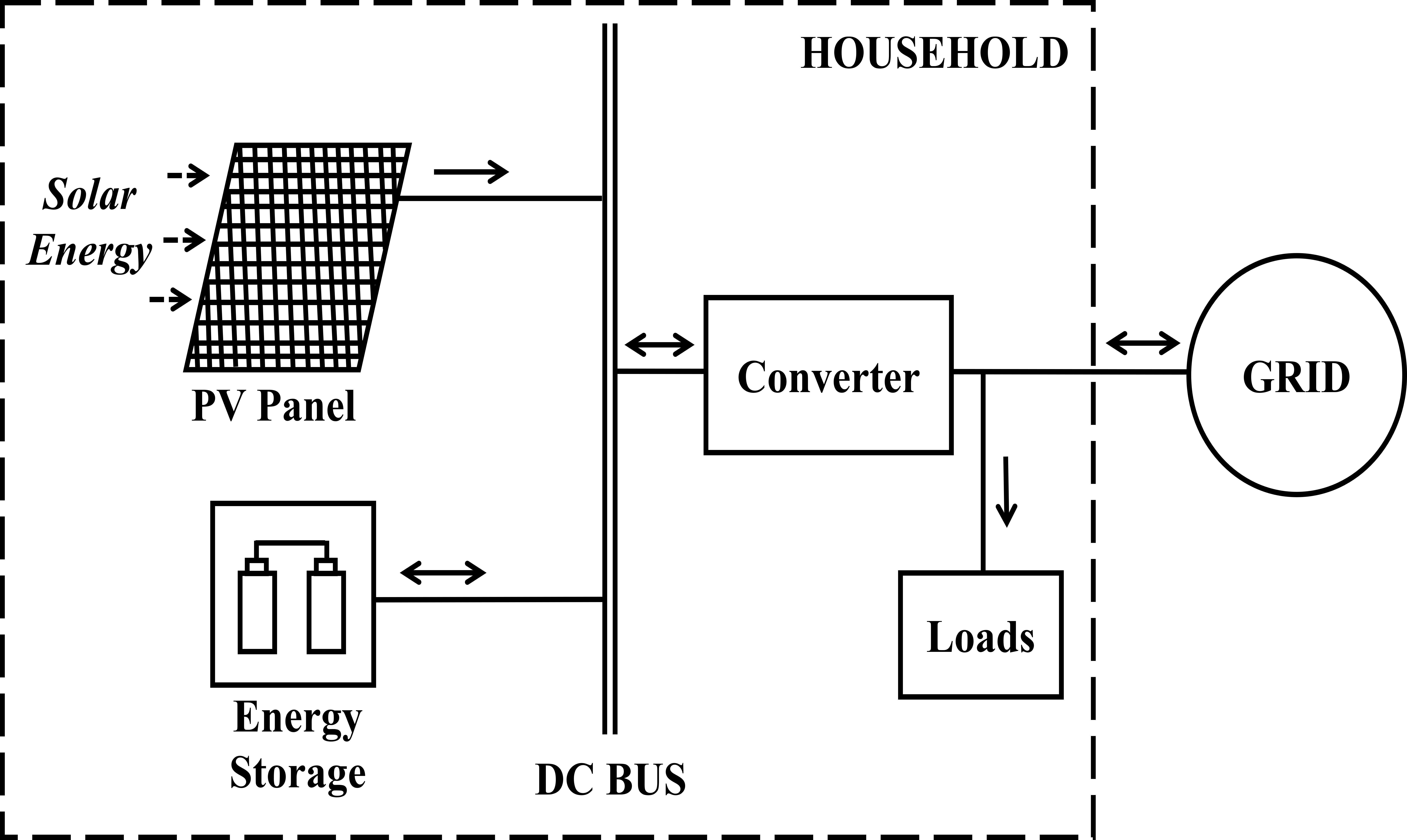}
  \caption{Block diagram of a single household with solar and storage interconnected to the grid.}
  \label{fig:blockdiagram}
\end{figure}

Optimal investment decisions in storage and solar panel for a household is formulated in this section. A single household with solar PV panel and energy storage device in grid-connected mode under normal loading conditions is shown in  Fig. \ref{fig:blockdiagram}.\\

The cost function of electricity consumption for the household is considered where each day is divided into two fixed continuous periods; peak ($h$) and off-peak ($l$) with ToU pricing and net metering billing mechanisms. The household electricity meter runs backward when excess power generation is fed to the grid under the net metering billing mechanism. At the end of a billing period, the house is compensated for the net power generation (i.e., net excess power fed by the household to the grid) at price $\mu$. Otherwise, the house would be required to pay the net consumption at price $\lambda$ for the deficit power consumed from the grid.

Various countries utilize diverse pricing systems for the promotion of renewable energy generation \cite{25}. Most states in the USA use a net metering system with $\lambda = \mu$. Japan employs NPS with $\mu \ge \lambda$. If the selling price is higher than the buying price i.e., $\mu > \lambda$, it would make solar panels and energy storage units very attractive for prosumers, creating significant electricity bill reductions. But this policy does not favour the utility companies as it does not take into account the transmission and distribution expenses which are covered by the utilities. And also the utilities would be required to purchase the generated power by households for a higher cost than the market cost which is not profitable. Therefore, in future we would have $\mu < \lambda$.

\noindent In our set-up, following pricing condition is considered: $\lambda_{h} > \mu_{h}  >\lambda_{l} > \mu_{l}$ where $\lambda_{h}$ , $\mu_{h}$, $\lambda_{l}$, $\mu_{l}$ are peak period buying price, peak period selling price, off-peak period buying price and off-peak period selling price respectively. For a random variable X, its expectation is written as $\mathbb{E}$[X]. The consumption of the household during peak and off-peak hours of a day are the random processes denoted by variables $H_h$ and $H_l$ respectively. Let $F_{H_h}(x)= \int_{-\infty}^{x} f_{H_h}(x)\,dx$ and $F_{H_l}(x)=\int_{-\infty}^{x} f_{H_l}(x)\,dx$ be the cumulative distribution functions, where $f_{H_h}(x)$ and $f_{H_l}(x)$ are the probability density function of $H_h$ and $H_l$ respectively for any day. The daily expected electricity consumption cost of the household (without solar and storage) with time-of-use pricing is
\begin {equation} \label{eq:costwithoutSandB}
C=\lambda_{h}\mathbb{E}[H_h] +\lambda_{l}\mathbb{E}[H_l]
\vspace{-2mm}
\end {equation}

\subsection{Optimal investment in storage capacity}
We first consider a household looking to invest in an energy storage capacity $B$. Let $\lambda_{b}$ be the daily capital cost of storage amortized over its life span. As the selling price during the peak period is greater than buying price during the off-peak period, i.e., $\mu_{h} > \lambda_{l}$, we can discharge the entire storage capacity during the peak period regardless of the consumption levels of the household. Thus the entire storage capacity is charged in the off-peak period.  We assume our storage device satisfies the necessary condition to have viable arbitrage opportunity, i.e., $(\mu_{h} - \lambda_{l}) \ge \lambda_{b}$. The household will now utilize the storage energy $(B)$ first and then purchase any deficit $(H_h - B)$ from the utility at the peak period price $\lambda_{h}$. When storage capacity is greater than the household consumption, the excess storage energy $(B-H_h)$ is sold back to the utility  at the peak period price $\mu_{h}$. During the off-peak period, the storage is only charged, and it is always charged to its full capacity. The household will purchase all the energy demand $(H_l+B)$ from the utility  at a lower price $\lambda_{l}$. The operating conditions of the storage unit are considered to be ideal. The daily expected cost of the household is
\begin{equation}\label{eq:costwithB}
    J(B)=\lambda_{b}B+\mathbb{E}[\lambda_{h}(H_h-B)^+-\mu_{h}(B-H_h)^+ + \lambda_{l}(H_l+B)]
\end{equation}
where $(x)^+=max\{x,0\}$ for any real number $x$. The household will invest in an optimal storage capacity,
\begin{equation*}
    B^0 = \argmin \; J(B)
\end{equation*}

\begin{thm}
The optimal investment decision of a household considering to purchase $B^0$ units of storage is

\begin{equation}\label{eq:F(B)}
    F_{H_h}(B^0)=\frac {\lambda_{h}-\lambda_{l}-\lambda_{b}} {\lambda_{h}-\mu_{h}}
\end{equation}

\noindent The optimal cost with storage investment is

\vspace{-4mm}
\begin{equation}\label{eq:costwithoptB}
J(B^0)=\lambda_{h}\mathbb{E}[H_h | H_h \ge B^0] + \mu_{h}\mathbb{E}[H_h | H_h < B^0]+\lambda_{l}\mathbb{E}[H_l]
\end{equation}
\end{thm}

\begin{proof}
$J(B)$ can be written as follows

\vspace{-2mm}
\begin{multline*}
        J(B)=\lambda_{b}B+\lambda_{h}\int_{B}^{\infty} (H_h-B)f_{H_h}(x)\,dx
        \\-\mu_{h}\int_{-\infty}^{B} (B-H_h)f_{H_h}(x) \,dx
        +\lambda_{l}\int_{-\infty}^{\infty} (H_l+B)f_{H_l}(x) \,dx
\end{multline*}

\noindent The optimal solution for $B^0$ using first order conditions is,

\begin{multline*}
       \frac{dJ(B)}{dB}=0\Rightarrow \lambda_{b} +\frac{d}{dB}\left[\lambda_{h}\int_{B}^{\infty} (H_h-B)f_{H_h}(x) \,dx\right]
       \\ -\frac{d}{dB}\left[\mu_{h}\int_{-\infty}^{B} (B-H_h)f_{H_h}(x) \,dx\right]
       \\ +\frac{d}{dB}\left[\lambda_{l}\int_{-\infty}^{\infty} (H_l+B)f_{H_l}(x) \,dx\right]
\end{multline*}

\noindent Using Leibnitz rule for differentiating integrals,

\begin{equation*}
\begin{split}
      0 =\lambda_{b}+\lambda_{h}\int_{B}^{\infty} (-1)f_{H_h}(x) &\,dx\ -\mu_{h}\int_{-\infty}^{B} (1)f_{H_h}(x) \,dx \\& +\lambda_{l}\int_{-\infty}^{\infty} (1)f_{H_l}(x) \,dx
\end{split}
\end{equation*}
\begin{equation*}
       =\lambda_{b}+\lambda_{h}(-1)[1-F_{H_h}(B)] -\mu_{h}F_{H_h}(B) +\lambda_{l}(1)
\end{equation*}

\noindent Therefore,

\begin{equation*}
    F_{H_h}(B^0)=\frac{\lambda_{h}-\lambda_{l}-\lambda_{b}}{\lambda_{h}-\mu_{h}}
\end{equation*}


\noindent Putting $B^0$ in the expression of $J(B)$ and simplifying, we get 

\begin{equation*}
    J(B^0)=\lambda_{h}\mathbb{E}[H_h | H_h \ge B^0] + \mu_{h}\mathbb{E}[H_h | H_h < B^0] + \lambda_{l}\mathbb{E}[H_l]
\end{equation*}
\end{proof}

Thus the optimal investment decision in storage capacity $B^0$ is dependent on the pricing conditions set by the utility and the daily capital cost of storage. The value of $B^0$ will be high depending on the following three conditions: if the difference in buying price from grid during peak and off-peak period ($\lambda_{h}-\lambda_{l}$) is high, if the daily capital cost of storage ($\lambda_{b}$) is low, and if the difference in the selling and buying price during peak-period ($\lambda_{h}-\mu_{h})$ is low. Thus the pricing conditions play a major role in obtaining the optimal storage capacity and also influences the storage capital costs as higher the value of $B^0$, higher will be the capital cost. It is to be noted that the battery's power ratings are not relevant for the proposed model as the charging period is quite long, and standard chargers used for battery charging can comfortably charge the battery to its full capacity in the off-peak period.

\subsection{Optimal investment in solar PV panel area}
Next we consider that the household has already invested in optimal energy storage of fixed capacity $B^0$ and is now looking to invest in solar PV panels with an area, $a$ optimally.
Let $S_h$ and $S_l$ be the solar irradiance level of the household during peak and off-peak periods, respectively. The consumption and irradiance levels of the household during peak and off-peak hours of a day are the random processes $H_h$, $H_l$ and $S_h$, $S_l$, respectively.
The household utilizes the storage and solar energy first and then purchases any deficit $(H_h-aS_h-B)$ from the utility at the peak period price $\lambda_{h}$. When storage capacity and solar generation are greater than the household consumption, the excess storage and solar energy $(aS_h+B-H_h)$ is sold back to the utility at the peak period price $\mu_{h}$. During the off-peak period, the storage is only charged to its full capacity, and the household looks to utilize only solar energy first and then purchases any deficit $(H_l+B-aS_l)$ at a lower price $\lambda_{l}$. When solar generation is greater than the household consumption, the excess solar energy $(aS_l-B-H_l)$ is sold back to the utility at a lower off-peak period price $\mu_{l}$. Let $\lambda_{a}$ be the daily capital cost of panel area amortized over its life span.
The operating conditions of the PV panels are considered to be ideal. The daily expected cost of the household is therefore
\begin{multline}\label{eq:costwithSandB}
    J(a)=\lambda_{b}B^0+\lambda_{a}a+\mathbb{E}[\lambda_{h}(H_h-aS_h-B)^+\\-\mu_{h}(aS_h+B-H_h)^+  +\lambda_{l}(H_l+B-aS_l)^+\\-\mu_{l}(aS_l-B-H_l)^+]
\end{multline}

\noindent The household will invest in a solar panel area,
\begin{equation*}
    a^0 = \argmin \; J(a) \quad \textrm{subject to}\quad 0 \leq a \leq a_{max}
\end{equation*}

\begin{thm}
The optimal investment decision of a household which already has an optimal storage unit installed and is considering to invest in $a^0$ area of solar PV panel under the condition $\lambda_h=\mu_h$ and $\lambda_l=\mu_l$ is,
\begin{equation} \label{eq:a^0}
a^0=\left\{
    \begin{aligned}
        & a_{max} \quad \textup{if} \quad  \lambda_{h}\mathbb{E}[S_h]+\lambda_{l}\mathbb{E}[S_l]\geq \lambda_{a}\\
        & 0 \quad \quad \; \; \textup{else}
    \end{aligned}\right.
\end{equation}
\end{thm}

\begin{proof}
If the buying and selling price for both peak and off-peak periods are equal, i.e., $\lambda_h=\mu_h$ and $\lambda_l=\mu_l$, the cost function $J(a)$ simplifies to
\begin{multline*}
    J(a)=\lambda_{b}B^0+\lambda_{a}a+\mathbb{E}[\lambda_{h}(H_h-aS_h-B) \\+\lambda_{l}(H_l+B-aS_l)]
\end{multline*}

The two inequality constraints are $-a$ $\le$ $0$ and $a-a_{max}$ $\leq$ $0$, and thus the Lagrangian function can be written as
\begin{multline*}
    L=\lambda_{b}B^0+\lambda_{a}a+\mathbb{E}[\lambda_{h}(H_h-aS_h-B) \\+\lambda_{l}(H_l+B-aS_l)]+\mu_1(-a)+\mu_2(a-a_{max})
\end{multline*}
The necessary Karush–Kuhn–Tucker conditions for finding the optimal condition of $a^0$ are
\begin{flalign*}
    & (i)\quad \frac{dL}{da}=0\Rightarrow\lambda_{a} -\lambda_{h}\mathbb{E}[S_h] -\lambda_{l}\mathbb{E}[S_l]-\mu_1+\mu_2=0, &\\
    & (ii)\quad \mu_1(-a)=0,  \quad (iii)\quad \mu_2(a-a_{max})=0,&\\
    & (iv)\quad -a\le0, \quad (v)\quad a-a_{max}\leq0,&\\
    & (vi)\quad \mu_1\ge0, \quad (vii)\quad \mu_2\geq0,&
\end{flalign*}
Solving based on these conditions, the optimal value of $a^0$ is given by equation (\ref{eq:a^0}),
\end{proof}

For the case of lower selling price for both peak and off-peak periods, i.e., $\lambda_h>\mu_h$ and $\lambda_l>\mu_l$ that we considered for the problem, though finding an analytical solution seems to be harder, we can still observe that the cost reduces with increase in $a$ under a condition that is a function of $\mathbb{E}[S_h],\mathbb{E}[S_l]$ along with the other prices including $\lambda_a$. But numerical methods are needed to find this condition.

Thus the optimal investment decision is to either invest in solar PV system with maximum panel area or not to invest at all. If the household has good average solar irradiance levels  ($\mathbb{E}[S_h],\mathbb{E}[S_l]$), then investing in  solar PV system with maximum possible panel area will lead to optimal reduction in the electricity cost. The maximum solar panel area is decided based on the availability of installation space and other factors such as budget, load demand, etc. On the other hand, if there is poor average solar irradiance levels throughout the year, then solar power generation will be low and investing in solar PV system will not help in reducing the electricity cost.  The result is in line with our intuition.

\begin{remark}
We have developed a model that is designed to be simple enough to be mathematically tractable for analysis and have ignored the aspects of complex pricing mechanisms, storage efficiency, storage device or solar cell degradation, operating costs etc. In reality, there can be more than two time periods with different prices, storage and solar devices can also degrade, storage devices have different charging and discharging efficiencies. There can also be operating cost. We can consider these conditions separately and arrive at analytical solutions for some of them, but we need a numerical simulation-based approach for analyzing the model for all conditions.
\end{remark}

\begin{table}[b]
  \centering
  \caption{Statistical Summary of Daily Average of Solar Irradiance and Daily Total of Load Consumption}
         \scalebox{0.75}
     {
    \begin{tabular}{>{\centering}m{4.055em} c c c c}
    \toprule
    \multicolumn{1}{r}{} & \multicolumn{1}{>{\centering}m{4.055em}}{Daily avg. solar irr. peak period (W/m\textsuperscript{2})} & \multicolumn{1}{>{\centering}m{4.055em}}{Daily total load peak period (kWh)} & \multicolumn{1}{>{\centering}m{4.055em}}{Daily avg. solar irr. off-peak period (W/m\textsuperscript{2})} & \multicolumn{1}{>{\centering}m{4.055em}}{Daily total load off-peak period (kWh)} \\
    \midrule
    \multicolumn{1}{c}{mean}  & 324.74 & 19.61 & 2.32  & 12.56 \\
    \multicolumn{1}{c}{std}  & 122.21 & 8.40  & 2.02  & 6.32 \\
    \multicolumn{1}{c}{min}   & 25.98 & 4.83 & -0.2  & 3.44 \\
    \multicolumn{1}{c}{25\%} & 254.73 & 12.90 & 0.77  & 7.05 \\
    \multicolumn{1}{c}{50\%} & 363.65 & 18.64 & 1.68  & 10.99 \\
    \multicolumn{1}{c}{75\%} & 418.21 & 24.79 & 3.44  & 17.81 \\
    \multicolumn{1}{c}{max}   & 505.85 & 51.42 & 10.01 & 34.32 \\
    \bottomrule
    \end{tabular}%
  \label{tab:stat}%
  \vspace{-4mm}
  }
\end{table}%

\section{Simulation Study and Result Analysis}
\label{section:Simulation Study and Result Analysis}

We consider a single household located in a residential area of Austin, Texas, that is looking to invest in PV rooftop panels with a storage unit. The data is taken from the Pecan Street project \cite{26}. The prosumer code of the single household selected for this study is 26. The period under study is the complete year of 2016. The average monthly consumption of this household is 1000 kWh. We divide the household consumption and the solar irradiance data into peak and off-peak periods.  We have considered peak-period from 8 hrs to 22 hrs and off-peak period from 22 hrs to 8 hrs. Table \ref{tab:stat} shows the statistical summary of the daily average of solar irradiance and daily total of load consumption for the household with both peak and off-peak periods. We observe that mean of the daily average of irradiance level is 324.74 W/m\textsuperscript{2} and mean of the daily total of household consumption is 19.61 kWh for peak period, and 2.32 W/m\textsuperscript{2} and 12.56 kWh for off-peak period, respectively. The standard rooftop solar PV panel produces 183 W at an irradiance of 1000 W/m\textsuperscript{2} for a panel area of 1 m\textsuperscript{2}. The PV panel output reduces to 170 W when considering 93\% PV system efficiency. We consider that the utility has set the buying price for peak and off-peak periods as 54 \mbox{\textcentoldstyle}/kWh and 22 \mbox{\textcentoldstyle}/kWh, and the selling price for peak and off-peak periods as 30 \mbox{\textcentoldstyle}/kWh and 13 \mbox{\textcentoldstyle}/kWh, respectively. These prices are inspired from the electricity prices of different US states\cite{22}, \cite{32}. The solar PV panel installation costs a national average of \$16,500 for a 6 kW solar panel system for a 1,500 square ft. home \cite{29}. We have considered the total capital cost for installing a solar PV panel in a residential household as 3 \$/W  which includes panel cost, mounting cost, labour cost, inverter cost, etc. \cite{29}. Thus, for a PV panel lifespan of 25 years, the amortized cost of solar panels per day is 5.58 \mbox{\textcentoldstyle}/m\textsuperscript{2}. Also, the total capital cost for installing a storage unit in a residential household is around 323 \$/kWh. Therefore, for a battery lifespan of 10 years, the amortized cost of storage unit per day is 8.84 \mbox{\textcentoldstyle}/kWh.

 \begin{figure}[]
  \centering
  \includegraphics[width=2.5in]{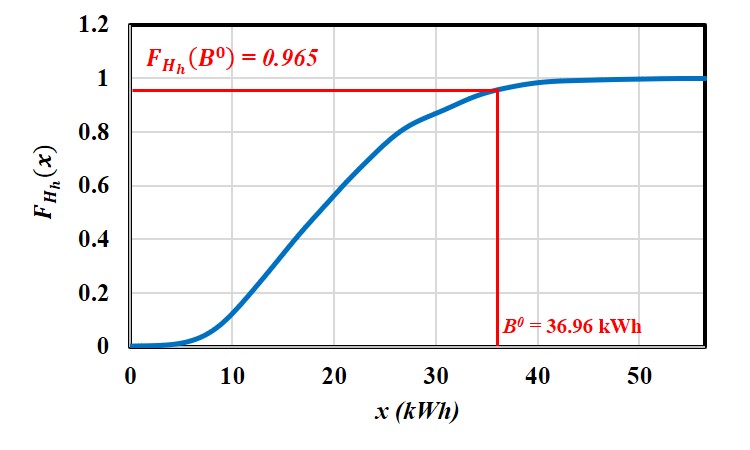}
  \caption{Cdf of peak-period daily total load consumption.}
  \label{fig:cdf}
  \vspace{-6mm}
\end{figure}
\begin{table*}[t]
    \centering
     \caption{Cost Analysis}
     \scalebox{0.675}
     {
    \begin{tabular}{>{\centering}m{4.055em} c c c c c c c c c c c c c c}
    \toprule
    \multicolumn{1}{c}{Cost $(\$)$} &\multicolumn{1}{>{\centering}m{4.055em}}{No B \& No PV} & \multicolumn{1}{>{\centering}m{4.055em}}{with only B\textsuperscript{0} \& no PV} & \multicolumn{1}{>{\centering}m{4.055em}}{with B\textsuperscript{0} \& PV of a=1m\textsuperscript{2}} & \multicolumn{1}{>{\centering}m{5.055em}}{with B\textsuperscript{0} \& PV of a=1.64m\textsuperscript{2}} & \multicolumn{1}{>{\centering}m{4.055em}}{with B\textsuperscript{0} \& PV of a=3m\textsuperscript{2}} & \multicolumn{1}{>{\centering}m{4.055em}}{with B\textsuperscript{0} \& PV of a=6m\textsuperscript{2}} & \multicolumn{1}{>{\centering}m{5.055em}}{with of B\textsuperscript{0} \& PV of a=12m\textsuperscript{2}} & \multicolumn{1}{>{\centering}m{5.055em}}{with B\textsuperscript{0} \& PV of a=15m\textsuperscript{2}} & \multicolumn{1}{>{\centering}m{5.055em}}{with of B\textsuperscript{0} \& PV of a=18m\textsuperscript{2}} & \multicolumn{1}{>{\centering}m{5.055em}}{with of B\textsuperscript{0} \& PV of a=24m\textsuperscript{2}} & \multicolumn{1}{>{\centering}m{5.055em}}{with of B\textsuperscript{0} \& PV of a=30m\textsuperscript{2}}\\
    \midrule
    \multicolumn{1}{c}{January} & 391.41  & 306.40  & 303.66  & 301.91  & 298.19  & 289.99  & 273.58  & 265.38  & 257.45  & 241.82  & 226.19  \\
    \multicolumn{1}{c}{February}  & 286.67  & 221.78  & 218.29  & 216.06  & 211.31  & 200.85  & 179.92  & 169.45  & 158.99  & 138.77  & 118.69  \\
    \multicolumn{1}{c}{March}  & 289.98  & 216.39  & 212.83  & 210.55  & 205.71  & 195.04  & 173.69  & 163.02  & 152.34  & 131.00  & 109.65  \\
    \multicolumn{1}{c}{April} & 300.38  & 223.57  & 219.95  & 217.64  & 212.72  & 201.88  & 180.18  & 169.34  & 158.49  & 136.80  & 115.11  \\
    \multicolumn{1}{c}{May} & 276.87  & 199.34  & 195.54  & 193.11  & 187.94  & 176.54  & 153.74  & 142.34  & 130.94  & 108.13  & 85.33   \\
    \multicolumn{1}{c}{June} & 339.77  & 248.96  & 244.35  & 241.39  & 235.12  & 221.28  & 193.59  & 179.75  & 165.91  & 138.23  & 110.54  \\
    \multicolumn{1}{c}{July} & 406.16  & 300.71  & 295.86  & 292.75  & 286.15  & 271.59  & 242.47  & 227.91  & 213.35  & 184.24  & 155.12  \\
    \multicolumn{1}{c}{August} & 350.13  & 256.86  & 252.66  & 249.97  & 244.25  & 231.64  & 206.41  & 193.80  & 181.19  & 155.97  & 130.74  \\
    \multicolumn{1}{c}{September} & 349.62  & 259.25  & 255.37  & 252.88  & 247.59  & 235.93  & 212.60  & 200.93  & 189.27  & 165.94  & 142.62  \\
    \multicolumn{1}{c}{October} & 300.12  & 217.40  & 213.17  & 210.45  & 204.69  & 191.98  & 166.55  & 153.84  & 141.13  & 115.70  & 90.28   \\
    \multicolumn{1}{c}{November}  & 379.40  & 281.79  & 278.55  & 276.47  & 272.06  & 262.34  & 242.89  & 233.16  & 223.44  & 203.99  & 185.01  \\
    \multicolumn{1}{c}{December}  & 350.99  & 261.05  & 258.63  & 257.08  & 253.79  & 246.52  & 231.99  & 224.72  & 217.46  & 202.93  & 188.40  \\
    \multicolumn{1}{c}{Total cost for one year} & 4021.49 & 2993.50 & 2948.84 & 2920.26 & 2859.53 & 2725.56 & 2457.62 & 2323.65 & 2189.96 & 1923.51 & 1657.67 \\
    \multicolumn{1}{c}{Capital cost for one year} & 0.00    & 1193.80 & 1214.20 & 1248.71 & 1255.00 & 1316.20 & 1438.60 & 1499.80 & 1561.00 & 1683.40 & 1805.80 \\
    \multicolumn{1}{c}{\begin{tabular}[c]{@{}c@{}} Total cost for one year \\ without capital cost\end{tabular}} & 4021.49    & 1799.70 & 1734.64 & 1671.56 & 1604.53 & 1409.36 & 1019.02 & 823.85  & 628.96  & 240.11  & -148.13 \\
    \multicolumn{1}{c}{\begin{tabular}[c]{@{}c@{}} Cost savings for one year \\ excluding capital cost\end{tabular}} &0.00 & 2221.79 & 2286.85 & 2349.93 & 2416.96 & 2612.13 & 3002.47 & 3197.64 & 3392.53 & 3781.38 & 4169.62 \\
    \multicolumn{1}{c}{\begin{tabular}[c]{@{}c@{}}Cost savings for one year \\ including capital cost\end{tabular}} & 0.00 & 1027.99 & 1072.65 & 1101.23 & 1161.96 & 1295.93 & 1563.87 & 1697.84 & 1831.53 & 2097.98 & 2363.82 \\
    \bottomrule
    \end{tabular}%
    }
  \label{tab:ana}%
  \vspace{-5mm}
\end{table*}%

 We first compute the optimal storage capacity $B^0$ for the household, which can be obtained by considering the daily peak-period total load consumption of each day for the entire year so that we have all possible values to compute $\mathbb{E}[H_h]$. We plot the cumulative distribution function (cdf) of the daily peak-period total load consumption using the kernel density estimation method \cite{27}. The value of $F_{H_h}(B^0)$ is calculated to be 0.965 from equation (\ref{eq:F(B)}), this value corresponds to the optimal value for the storage capacity $B^0$ on the cdf curve of $F_{H_h}(x)$ shown in Fig. \ref{fig:cdf}, the value of $B^0$ is found to be 36.96 kWh.

The simulations are carried out using the data of each day in the year. As there are 366 days in the year 2016, we simulate for all 366 days and then calculate the monthly cost of the household. The solar panel area of around 6 m\textsuperscript{2} roughly amounts to 1 kWp, so we compare the cost difference for each kilo-watt upto 5 kWp. In the US, the average household roof area is 12 to 15 m\textsuperscript{2}. For a panel area of 15 m\textsuperscript{2}, the solar panel output capacity will be 2.5 kWp. We have analysed upto 5 kWp with maximum panel area of 30 m\textsuperscript{2}. A more detailed cost analysis is provided in Table \ref{tab:ana}. The total expected cost of the household without storage and without solar is \$4,021.49. After investing in optimal storage capacity $B^0$, the cost reduces to \$2,993.50 with 25.56\% in savings including the capital cost for one year. There is an additional reduction of \$134 in the total cost with 3.3\% savings for every 0.5 kW of solar PV panel investment with panel area of 3 m\textsuperscript{2}. Thus we can observe that though the capital cost of storage is high, the cost savings cover the capital cost and also the investment in solar provides additional savings depending on the amount of investment capacity or area.

But we understand that if a prosumer wants to invest in solar PV and storage units simultaneously, this approach will give a sub-optimal solution. Around the neighborhood of ($B^0$, $a^0$), other points are found where the total cost is reduced. For example, for $a^0$ = 30, the total cost is \$1657.67 for $B^0$ = 37. But the total cost reduces to \$1612.78 for $B$ = 24.

\section{Conclusion}
\label{section:Conclusion}
In this paper, we studied the impact of optimal investment decisions for storage and solar PV panels in a single household under net metering billing mechanism with time-of-use pricing. We presented a case study using load consumption and solar irradiance data of one year, and investigated how optimal investment decisions affect the net electricity consumption cost of the household. We showed that through optimal investments in storage and solar, significant cost benefits can be attained for the household. This work provides valuable information for the prosumer to consider before investing in solar and storage resources under the new billing mechanism.

In future, we plan to derive joint investment decision of a prosumer who would invest in both solar PV and storage unit simultaneously instead of investing on one after another. We also plan to consider scenarios like more prices in a day, price uncertainties, battery degradation etc. in our analysis.

\bibliographystyle{IEEEtran}
\bibliography{LatexTemplate/main}

\end{document}